\documentstyle[aaspp4,flushrt]{article}

\makeatletter
\def\thebibliography#1{\section*{REFERENCES}
 \addcontentsline{toc}{section}{REFERENCES}
 \list{}{\labelwidth\z@
         \leftmargin 1.5em
	 \itemsep \z@
	 \itemindent-\leftmargin}
 \small\raggedright
 \parindent\z@
 \parskip\z@ plus .1pt\relax
 \def\newblock{\hskip .11em plus .33em minus .07em}
 \sloppy\clubpenalty4000\widowpenalty4000
 \sfcode`\.=1000\relax
}

\def\@biblabel#1{}
\def\@bcite#1#2{(#1\if@tempswa , #2\fi)}
\def\@pcite#1#2{#1\if@tempswa , #2\fi}
\def\@citefmta#1#2{#1 (#2)}
\def\@citefmtb#1#2{#1 #2}
\let\citefmt=\@citefmta
\def\@citex[#1]#2{\if@filesw\immediate\write\@auxout{\string\citation{#2}}\fi
  \def\@citea{}\@cite{\@for\@citeb:=#2\do
    {\@citea\def\@citea{;\penalty\@m\ }\@ifundefined
    {b@\@citeb}{{\bf ?}\@warning
{Citation `\@citeb' on page \thepage \space undefined}}%
{\csname b@\@citeb\endcsname}}}{#1}}
\def\cite{\@ifnextchar [{\let\citefmt=\@citefmtb
                          \let\@cite=\@bcite\@tempswatrue \@citex}
                        {\let\citefmt=\@citefmtb
                          \let\@cite=\@bcite\@tempswafalse \@citex[]}}
\def\pcite{\@ifnextchar [{\let\citefmt=\@citefmtb
                          \let\@cite=\@pcite\@tempswatrue\@citex}
                        {\let\citefmt=\@citefmtb
                          \let\@cite=\@pcite\@tempswafalse\@citex[]}}
\def\scite{\@ifnextchar [{\let\citefmt=\@citefmta
                          \let\@cite=\@pcite\@tempswatrue\@citex}
                        {\let\citefmt=\@citefmta
                          \let\@cite=\@pcite\@tempswafalse\@citex[]}}
\makeatother

\def\binko#1#2{\bigg(\!\begin{array}{c}{#1}\\[-1ex]{#2}\end{array}\!\bigg)}
\def\tfrac#1#2{{\textstyle\frac{#1}{#2}}}
\def\dfrac#1#2{{\displaystyle\frac{#1}{#2}}}

\def\ddx{{{\rm d}^d\!x\,}}
\def\dt{{{\rm d}t}}
\def\dx{{{\rm d}x}}

\def\bx{{\bf x}}
\def\cD{{\cal D}}
\def\cE{{\cal E}}
\def\cL{{\cal L}}
\def\rd{{\rm d}}
\def\rv{{\rm v}}
\def\RR{{\rm I\kern-0.16em{}R}}

\begin{document}
\title{Beyond genus statistics:\\
a unifying approach to the morphology of cosmic structure}
\author{Jens Schmalzing\altaffilmark{1,2,3} and Thomas
Buchert\altaffilmark{1,4}}
\affil{Ludwig--Maximilians--Universit\"at M\"unchen}
\altaffiltext{1}{Ludwig--Maximilians--Universit\"at, Theresienstra{\ss}e 37,
80333 M\"unchen, Germany}
\altaffiltext{2}{Max--Planck--Institut f\"ur Astrophysik,
Karl--Schwarzschild--Stra{\ss}e 1, 85740 Garching, Germany}
\altaffiltext{3}{email jensen@mpa-garching.mpg.de}
\altaffiltext{4}{email buchert@stat.physik.uni-muenchen.de}

\begin{abstract}
\noindent 
The genus statistics of isodensity contours has become a
well--established tool in cosmology. In this {\em Letter} we place
the genus in the wider framework of a complete family of
morphological descriptors. These are known as the Minkowski
functionals, and we here apply them for the first time to isodensity
contours of a continuous random field. By taking two equivalent
approaches, one through differential geometry, the other through
integral geometry, we derive two complementary formulae suitable for
numerically calculating the Minkowski functionals. As an example we
apply them to simulated Gaussian random fields and compare the outcome
to the analytically known results, demonstrating that both are indeed
well suited for numerical evaluation. The code used for calculating
all Minkowski functionals is available from the authors.
\end{abstract}


\section{Introduction}

The ambitious goal of characterizing the morphology of cosmic
structure in a unique and complete way leads to the field of integral
geometry. There, this problem has been well--posed and Hadwiger's
theorem (\pcite{hadwiger:vorlesung}, see also {}\pcite{klain:hadwiger}
for a proof) gives a powerful answer: In $d$ spatial dimensions the
global morphological properties (defined as those which satisfy
motional invariance and additivity) of any pattern can be completely
characterized by $d+1$ numbers, the so--called Minkowski functionals
{}\cite{minkowski:volumen}. Furthermore, in the interesting cases of
two and three dimensions, all the Minkowski functionals can be
intuitively interpreted in terms of well--known geometric quantities.

One of them is the Euler characteristic $\chi$, or equivalently the
genus $g=1-\chi$, which has long been used as a statistical tool in
cosmology. The standard analysis technique starts from a density or
temperature field. For a point process it is usually constructed by
applying a smoothing filter with a constant width larger than the mean
separation of points. For this field, the Euler characteristic of an
isocontour at level $\nu$ is obtained, according to the Gauss--Bonnet
theorem, through a surface integration of the Gaussian curvature. This
threshold level can be employed as a diagnostic parameter, and the
genus represented as a function of $\nu$ (see
{}\pcite{weinberg:topologyI}, {}\pcite{melott:review},
{}\pcite{coles:quantifying} and references therein) can be compared
with analytically known values for Gaussian random fields
{}\cite{doroshkevich:gauss} and even for some nonlinear dynamical
models such as solutions of the Lagrangian perturbation theory
{}\cite{matsubara:evolution,seto:nonlinear}.

This method for calculating the genus or, as advocated in the present
{\em Letter}, all four Minkowski functionals, suffers from erasing
small--scale information {}\cite{koenderink:scalespace} when going
from point sets to smooth density fields. While this may actually be
an advantage when trying to deemphasize poorly understood effects such
as nonlinear evolution and biasing, some discriminatory power is lost.

In this situation, {}\scite{mecke:robust} introduced Minkowski
functionals of point processes into cosmology. Proceeding as directly
as possible they decorate each point with a ball of radius $R$, which
becomes the diagnostic parameter for this method of analysis. Thereby
a union set of balls is constructed and probes the morphological
properties of the point set. These are measured in a complete and
unique way using Minkowski functionals, so powerful methods of
integral geometry {}\cite{weil:stereology} become applicable.  Most
notably, Minkowski functionals contain a variety of other descriptors
such as the ``void probability function'' {}\cite{white:hierarchy},
and depend on the complete hierarchy of correlations
{}\cite{stratonovich:topicsI,mecke:robust}. Analytically known results
for the Poisson process {}\cite{mecke:euler} provide a useful standard
of reference; the additivity property in integral geometry leads to
robustness against various sources of error
{}\cite{kerscher:fluctuations}; the sensitivity to small--scale
clustering properties allows for efficient discrimination, as has been
demonstrated in recent applications to the Abell/ACO cluster catalogue
{}\cite{kerscher:abell} and the IRAS 1.2Jy catalogue
{}\cite{kerscher:fluctuations}. A brief tutorial on Minkowski
functionals in cosmology can be found in
{}\scite{schmalzing:minkowski}.

In this {\em Letter} we advocate the use of the whole family of
Minkowski functionals to analyze the continuous random fields
encountered in cosmology.  Thereby we obtain more information than by
measuring the genus alone. Furthermore, recent work
(\pcite{kerscher:fluctuations}, see also {}\pcite{buchert:geltow} for
a direct comparison) has shown that other Minkowski functionals such
as the surface area and the integrated mean curvature can possess
greater discriminatory power than the Euler characteristic. Apart from
this practical advantage, which was already pointed out for the area
by {}\scite{ryden:area}, the use of all the Minkowski functionals
leads to a conceptual generalization of genus statistics, incorporates
it into the firm mathematical framework of integral geometry and
stereology and allows for a straightforward evaluation, including
exact correction for boundary effects.

The {\em Letter} is organized as follows.  Section 2 uses the methods
of differential geometry to reduce the evaluation of Minkowski
functionals to a spatial average of Koenderink invariants. Section 3
employs integral geometry, namely Crofton's intersection formula, to
derive a second calculational method that relies on the representation
of the density field by its values at lattice points. Section 4 gives
the analytical formulae 
for all the Minkowski functionals of Gaussian random fields, while
Section 5 compares them to the results obtained by applying our two
approximation formulae to numerically simulated Gaussian random
fields. Section 6 provides an outlook.

\section{Koenderink Invariants}

Consider a random field $\nu(\bx)$ on a $d$--dimensional support
$\cD\subseteq\RR^d$.  For a given threshold $\nu$ the excursion set
$F_\nu$ is the set of all points $\bx$ with $\nu(\bx)\geq{\nu}$.  We
wish to calculate its Minkowski functionals $\rv_k^{(d)}(\nu)$ per
unit volume. The volume functional $\rv_0^{(d)}$ is simply given by
volume integration of a Heaviside step function $\Theta$, normalized
to the whole volume $|\cD|$,
\begin{equation}
\rv_0^{(d)}(\nu) = \dfrac{1}{|\cD|} \int_\cD\ddx \Theta(\nu-\nu(\bx)).
\end{equation}

In three dimensions, the other Minkowski functionals correspond to
quantities known from differential geometry, namely the surface area,
the integrated mean curvature and the Euler characteristic
{}\cite{schneider:brunn}. They can be evaluated by a surface
integration
\begin{equation}
\rv_k(\nu)=\dfrac{1}{|\cD|} \int_{\partial{F}_\nu} \rd^2\!\!A(\bx) \rv_k^{({\rm loc})}(\nu,\bx).
\label{eq:surface}
\end{equation}
The local Minkowski functionals
\begin{equation}
\begin{array}{rcl}
\rv_1^{({\rm loc})}(\nu,\bx)&=&\dfrac{1}{6} \\[1ex]
\rv_2^{({\rm loc})}(\nu,\bx)&=&\dfrac{1}{6\pi} \left(\dfrac{1}{R_1}+\dfrac{1}{R_2}\right)\\[1ex]
\rv_3^{({\rm loc})}(\nu,\bx)&=&\dfrac{1}{4\pi} \dfrac{1}{R_1R_2}
\end{array}
\end{equation}
are nothing but appropriately normalized invariants made from the
inverse radii of curvature $R_1$ and $R_2$ of the surface orientated
towards lower density values.

Using methods developed by Koenderink
{}\cite{koenderink:structure,koenderink:scalespace} in
two--dimensional image analysis, it is possible to express these local
curvatures in terms of geometric invariants formed from first and
second derivatives {}\cite{schmalzing:diplom}.  The surface
integration (\ref{eq:surface}) is equivalent to
taking the spatial mean over the whole volume, with a gradient for the
surface element and a delta function for
selecting the isodensity contour added as factors. The resulting
formula is
\begin{equation}
\rv_k(\nu)= \dfrac{1}{|\cD|} \int_\cD \rd^3\!x\, \delta(\nu-\nu(\bx))
\left|\nabla{\nu(\bx)}\right| \rv_k^{({\rm loc})}(\nu,\bx).
\label{eq:koenderink}
\end{equation}
If the density field is sampled at some lattice points, we can
therefore estimate the Minkowski functionals by calculating the
derivatives at each lattice point and replacing first the
$\delta$-function with a bin of finite width and the average over all
space with a sum over lattice points only.  Analogous formulae can be
derived in two dimensions, so the most promising cosmological
application of this particular method is an elegant way of calculating
Minkowski functionals for fields on the sphere, like all--sky maps of
the CMB (for existing work, see {}\pcite{smoot:topologyfirst},
{}\pcite{colley:topology} and especially {}\pcite{torres:genus}).

\section{Crofton's Intersection Formula}

Another powerful approach to morphology of density fields comes
through integral geometry
{}\cite{hadwiger:vorlesung,santalo:integralgeometry}. As above we
consider a random field $\nu(\bx)$ and measure the Minkowski functionals
$\rv_k^{(d)}(\nu)$ of its excursion set over a threshold $\nu$.

To do this we use Crofton's formula {}\cite{crofton:theory}.
For a body $K$ in $d$ dimensions we consider an arbitrary
$k$--dimensional hyperplane $E$ and calculate the Euler characteristic
$\chi$ of the intersection $K{\cap}E$ in $k$ dimensions.  Integrating
this quantity over the space $\cE_k^{(d)}$ of all conceivable
hyperplanes\footnote{The integration measure $\rd\mu_k(E)$ is
normalized to give $\int_{\cE_k^{(d)}}\rd\mu_k(E)=1$. } we obtain the
$k$th Minkowski functional of the body $K$ in $d$
dimensions. Crofton's formula states that\footnote{$\omega_k$ is the
volume of the unit ball in $k$ dimensions, so $\omega_0=1$,
$\omega_1=2$, $\omega_2=\pi$ and $\omega_3=\tfrac{4}{3}\pi$.}
\begin{equation}
\rv_k^{(d)}(K) = \dfrac{\omega_d}{\omega_{d-k}\omega_k}
\int_{\cE_k^{(d)}}\rd\mu_k(E) \chi^{(k)}(K{\cap}E).
\end{equation}
If the body $K$ is the excursion set of a homogeneous and isotropic
random field sampled at $L$ points of a cubic lattice of spacing $a$,
we can as well average over the set $\cL_k^{(d)}$ of all dual lattice
hyperplanes only. The integral is then replaced by a summation over
these hyperplanes with an additional normalization factor and we
obtain
\begin{equation}
\rv_k^{(d)}(K) = \dfrac{\omega_d}{\omega_{d-k}\omega_k}
\sum_{E\in\cL_k^{(d)}}\dfrac{1}{a^kL}\dfrac{k!(d-k)!}{d!}
\chi^{(k)}(K{\cap}E).
\end{equation}
However, the Euler characteristic of a body in $k$ dimensions can be
approximated by counting the numbers $N_j(K)$ of all elementary
lattice cells of dimension $j$ within the body
{}\cite{adler:randomfields}. In three dimensions, for example,
$N_3(K)$ gives the number of cubes within the body, $N_2(K)$ and
$N_1(K)$ are the numbers of faces and edges of the lattice cubes,
respectively, while $N_0(K)$ gives the number of lattice points
contained in $K$. In general
\begin{equation}
\chi^{(k)}(K{\cap}E) = \sum_{j=0}^k(-1)^jN_j(K{\cap}E).
\label{eq:euler}
\end{equation}
When summing equation (\ref{eq:euler}) over all lattice hyperplanes,
we notice that $j$--dimensional lattice cells of the bodies $K{\cap}E$
result from intersections of a $(d-k+j)$--dimensional lattice cell
belonging to the body $K$ with $k$--dimensional hyperplanes $E$, so
there are $\binko{d-k+j}{j}$ of them. So we can actually perform the
summation
\begin{equation}
\sum_{E\in\cL_k^{(d)}}N_j(K{\cap}E) =
\binko{d-k+j}{j}N_{d-k+j}(K)
\end{equation}
and end up with the compact formula
\begin{equation}
\rv_k^{(d)}(K) = \dfrac{\omega_d}{\omega_{d-k}\omega_k}
\dfrac{1}{a^kL}\sum_{j=0}^k(-1)^j\dfrac{k!(d-k+j)!}{d!j!}
N_{d-k+j}(K).
\end{equation}
For a homogeneous and isotropic random field in three dimensions the
Minkowski functionals are thus given by
\begin{equation}
\begin{array}{rcl}
\rv_0(K)&=&\dfrac{1}{L} N_3(K)\\[1ex]
\rv_1(K)&=&\dfrac{1}{aL}(-\tfrac{2}{3}N_3(K)+\tfrac{2}{9}N_2(K))\\[1ex]
\rv_2(K)&=&\dfrac{1}{a^2L}(\tfrac{2}{3}N_3(K)-\tfrac{4}{9}N_2(K)+\tfrac{2}{9}N_1(K))\\[1ex]
\rv_3(K)&=&\dfrac{1}{a^3L}  (-N_3(K)+ N_2(K)-N_1(K)+N_0(K) ).
\end{array}
\label{eq:crofton}
\end{equation}
Note that the last relation for the Euler characteristic
$\rv_3=\chi/|\cD|$ per unit volume is not obtained from Crofton's
formula but is already given as a special case of equation
(\ref{eq:euler}) which results from the Morse theorem
{}\cite{morse:critical}. Being conceptually simpler than the other
approximations, it was proved rigorously by
{}\scite{adler:randomfields} and used by {}\scite{coles:quantifying}
to calculate the genus of isodensity contours in three
dimensions. Adding Crofton's formula gives all the Minkowski
functionals without further computational effort, since the
calculation requires nothing but the counting of all elementary
lattice cells, which is already necessary for the genus.

\section{Tomita's Formulae for Statistics of Random Interfaces}

Like the genus {}\cite{doroshkevich:gauss}, all Minkowski functionals
are analytically known for a Gaussian random field $\nu(\bx)$ in $d$
dimensions, due to a highly instructive article by
{}\scite{tomita:statistics}. Again, consider the excursion set over a
threshold $\nu$ and calculate its mean Minkowski functionals
$\rv_k^{(d)}(\nu)$ per unit volume.

The result depends on only two parameters $\sigma$ and $\lambda$,
which in turn depend on the value of the correlation function $\xi(0)$
and its second derivative $\xi''(0)$ at zero through
\begin{equation}
\sigma=\xi(0),\qquad \lambda=\sqrt{|\xi''(0)|/(2\pi\xi(0))}.
\end{equation}
Both are easily measured from a given realization of the random field,
since $\xi(0)=\langle{\nu^2}\rangle$ is the variance of the field
itself and $|\xi''(0)|=\langle{\nu_{,i}^2}\rangle$ is the
variance of any of its first derivatives {}\cite{adler:randomfields}.

With these two parameters all Minkowski functionals in arbitrary
dimension $d$ can be calculated for a Gaussian random field
{}\cite{tomita:statistics}. For the volume functional we
obtain\footnote{The function
$\Phi(x)=\tfrac{2}{\sqrt{\pi}}\int_0^x\dt\exp(-t^2)$ denotes the
Gaussian error integral, while $H_n(x) =
\left(-\tfrac{\rd}{\dx}\right)^n \tfrac{1}{\sqrt{2\pi}}
\exp\left(-\tfrac{1}{2}x^2\right)$ defines a Hermite function of order
$n$.}
\begin{equation}
\rv_0^{(d)}(\nu) =
\tfrac{1}{2}-\tfrac{1}{2}\Phi\left(\dfrac{\nu}{\sqrt{2\sigma}}\right),
\end{equation}
while all other Minkowski functionals are given by
\begin{equation}
\rv_k^{(d)}(\nu) = \lambda^k \dfrac{\omega_d}{\omega_{d-k}\omega_k}
H_{k-1}\left(\dfrac{\nu}{\sqrt{\sigma}}\right).
\end{equation}
Putting $d=3$ we obtain the Minkowski functionals in three
dimensions\footnote{To relieve the notation of the overall
normalization $\sqrt{\sigma}$, we use the dimensionless argument $u$,
with $\nu=u\sqrt{\sigma}$.}
\def\expo{\exp\left(-\tfrac{1}{2}u^2\right)}
\begin{equation}
\begin{array}{rcl}
\rv_0(\nu)&=&\tfrac{1}{2}-\tfrac{1}{2}\Phi\left(\tfrac{1}{\sqrt{2}}u\right)\\[1ex]
\rv_1(\nu)&=&\tfrac{2}{3}\dfrac{\lambda}{\sqrt{2\pi}}\expo\\[1ex]
\rv_2(\nu)&=&\tfrac{2}{3}\dfrac{\lambda^2}{\sqrt{2\pi}} u \expo\\[1ex]
\rv_3(\nu)&=&\dfrac{\lambda^3}{\sqrt{2\pi}}\left(u^2-1\right)\expo.
\end{array}
\label{eq:tomita}
\end{equation}

\section{Gaussian Random Fields}

Gaussian random fields provide a useful standard of reference and are
always good for a pedagogical example. Here we use a scale--invariant
power spectrum to construct Gaussian random fields on $128^3$ points
in a unit cube. We apply a Gaussian smoothing filter of width
$16/128=0.125$, which is chosen large in order to give a large
variation between realizations. The field is normalized to
$\left\langle\nu\right\rangle=0$ and
$\left\langle\nu^2\right\rangle=1$. Then, we calculate the numerical
estimates and the analytical predictions at 100 threshold values
between $-4$ and $4$ and obtain the mean and variance over 200 such
curves.

Figure~\ref{fig:gauss} shows three areas in each plot. One is measured
by averaging Koenderink invariants as in equation
(\ref{eq:koenderink}), the second by using Crofton's formula
(\ref{eq:crofton}). However, these two are completely
indistinguishable, so the two approximations are indeed equivalent.

A third area (shaded, dotted lines) is obtained by estimating the
parameters $\sigma$ and $\lambda$ from each random field and computing
the analytical mean Minkowski functionals according to
(\ref{eq:tomita}). Obviously the measured mean values fit the
analytical expectations extremely well; deviations are far from
significant. Since the analytical curves incorporate only statistical
fluctuations in the determination of the parameters, their variance is
reduced compared to the measured curves, which contain the full
variance between and within samples. Obviously, surface area $\rv_1$
and integrated mean curvature $\rv_2$ vary almost exclusively through
parameter fluctuations, so these measures enhance the discriminatory
power compared to considering the Euler characteristic $\rv_3$ alone.

\section{Summary and Outlook}

We have extended the scope of genus statistics by considering all four
Minkowski functionals in three dimensions. Two independent approaches
gave us two different approximation formulae which we have tested by
application to a Gaussian random field. Furthermore, we verified that
our measurements fit the analytically known Minkowski functionals of a
Gaussian random field.

Even without analytical reference values, we have two complementary
numerical formulae at our disposal which both rely on the
approximation of a smooth density field by values at lattice
points. The agreement between the two can be employed to determine the
necessary smoothing length. Too little smoothing would be easily
detectable by discrepancies between the two formulae.

Forthcoming applications of the presented method will give intuitive
interpretations of our novel access to cosmic morphology. Practical
issues such as boundary corrections (see {}\pcite{kerscher:minkowski}
for point processes), and further tests of the robustness and
discriminatory power are necessary beyond the promising results given
here.


\section*{Acknowledgements}

We thank Claus Beisbart, Martin Kerscher, Adrian Melott, Herbert
Wagner, David Weinberg, and especially Simon White for interesting
discussions and valuable comments. TB acknowledges support by the
``Sonderforschungsbereich 375 f\"ur Astroteilchenphysik der Deut\-schen
For\-schungs\-ge\-mein\-schaft''.

\section*{}

The code for calculating all Minkowski functionals as functions of
threshold for a continuous random field sampled on a cubic lattice can
be obtained from the authors.

\newpage

\providecommand{\bysame}{\leavevmode\hbox to3em{\hrulefill}\thinspace}

\newpage

\section*{Figure Captions}
\begin{figure*}
\caption{\label{fig:gauss}
The four Minkowski functionals of isodensity contours of a Gaussian
random field in three dimensions. The central lines show the exact
mean values while the area between top and bottom lines indicates the
statistical variance of our results over 200 realizations. Actually
there are three areas in each plot, corresponding to the three
formulae (\protect\ref{eq:koenderink}), (\protect\ref{eq:crofton}) and
(\protect\ref{eq:tomita}). However, the two independent approximation
formulae (no shading) perform so well that their results are
identical. Also, the agreement with analytically predicted mean values
(shaded) is remarkable. }
\end{figure*}


\begin{thebibliography}{\protect\citefmt{ter Haar~Romeny \bgroup et al.\egroup
  }{1991}}

\bibitem[\protect\citefmt{Adler}{1981}]{adler:randomfields}
Adler, R.~J., \emph{The geometry of random fields}, John Wiley \& Sons,
  Chichester, 1981.

\bibitem[\protect\citefmt{Buchert}{1995}]{buchert:geltow}
Buchert, T., in: \emph{11th Potsdam Workshop on Large--Scale Structure in the
  Universe} (Geltow) (M{\"u}cket, J., Gottl{\"o}ber, S., \& M{\"u}ller, V.,
  eds.), World Scientific, 1995, pp.~156--161.

\bibitem[\protect\citefmt{Coles \bgroup et al.\egroup
  }{1996}]{coles:quantifying}
Coles, P., Davies, A., \& Pearson, R.~C., Mon. Not. R. Astron. Soc.
  \textbf{281} (1996), 1375--1384.

\bibitem[\protect\citefmt{Colley \bgroup et al.\egroup
  }{1996}]{colley:topology}
Colley, W.~N., {Gott III}, J.~R., \& Park, C., Mon. Not. R. Astron. Soc.
  \textbf{281} (1996), L82--L84.

\bibitem[\protect\citefmt{Crofton}{1868}]{crofton:theory}
Crofton, M.~W., Phil.\ Trans.\ Roy.\ Soc.\ London \textbf{158} (1868),
  181--199.

\bibitem[\protect\citefmt{Doroshkevich}{1970}]{doroshkevich:gauss}
Doroshkevich, A.~G., Astrophysics \textbf{6} (1970), 320--330.

\bibitem[\protect\citefmt{Hadwiger}{1957}]{hadwiger:vorlesung}
Hadwiger, H., \emph{Vorlesungen {\"u}ber Inhalt, Oberfl{\"a}che und
  Isoperimetrie}, Springer Verlag, Berlin, 1957.

\bibitem[\protect\citefmt{Kerscher \bgroup et al.\egroup
  }{1996}]{kerscher:minkowski}
Kerscher, M., Schmalzing, J., \& Buchert, T., in: \emph{Mapping, measuring and
  modelling the universe} (Valencia) (Coles, P., {Mart\'{\i}nez}, V., \& {Pons
  Border\'{\i}a}, M.~J., eds.), Astronomical Society of the Pacific, 1996,
  pp.~247--252.

\bibitem[\protect\citefmt{Kerscher \bgroup et al.\egroup
  }{1997}a]{kerscher:fluctuations}
Kerscher, M., Schmalzing, J., Buchert, T., \& Wagner, H., to be submitted,
  1997.

\bibitem[\protect\citefmt{Kerscher \bgroup et al.\egroup
  }{1997}b]{kerscher:abell}
Kerscher, M., Schmalzing, J., Retzlaff, J., Borgani, S., Buchert, T.,
  Gottl\"ober, S., M\"uller, V., Plionis, M., \& Wagner, H., Mon. Not. R.
  Astron. Soc. \textbf{284} (1997), 73--84.

\bibitem[\protect\citefmt{Klain}{1995}]{klain:hadwiger}
Klain, D.~A., Mathematika \textbf{42} (1995), 329--339.

\bibitem[\protect\citefmt{Koenderink}{1984}]{koenderink:structure}
Koenderink, J.~J., Biol. Cybern. \textbf{50} (1984), 363--370.

\bibitem[\protect\citefmt{Matsubara \& Suto}{1996}]{matsubara:evolution}
Matsubara, T. \& Suto, Y., Ap. J. \textbf{460} (1996), 51--58.

\bibitem[\protect\citefmt{Mecke \& Wagner}{1991}]{mecke:euler}
Mecke, K. \& Wagner, H., J. Stat. Phys. \textbf{64} (1991), 843.

\bibitem[\protect\citefmt{Mecke \bgroup et al.\egroup }{1994}]{mecke:robust}
Mecke, K., Buchert, T., \& Wagner, H., Astron. Astrophys. \textbf{288} (1994),
  697.

\bibitem[\protect\citefmt{Melott}{1990}]{melott:review}
Melott, A.~L., Physics Rep. \textbf{193} (1990), 1--39.

\bibitem[\protect\citefmt{Minkowski}{1903}]{minkowski:volumen}
Minkowski, H., Mathematische Annalen \textbf{57} (1903), 447--495.

\bibitem[\protect\citefmt{Morse \& Cairns}{1969}]{morse:critical}
Morse, M. \& Cairns, S.~S., \emph{Critical point theory in global analysis and
  differential topology}, Academic Press, New York and London, 1969.

\bibitem[\protect\citefmt{Ryden \bgroup et al.\egroup }{1989}]{ryden:area}
Ryden, B.~S., Melott, A.~L., Craig, D.~A., {Gott III}, J.~R., Weinberg, D.~H.,
  Scherrer, R.~J., Bhavsar, S.~P., \& Miller, J.~M., Ap. J. \textbf{340}
  (1989), 647--660.

\bibitem[\protect\citefmt{Santal\'o}{1976}]{santalo:integralgeometry}
Santal\'o, L.~A., \emph{Integral Geometry and Geometric Probability},
  Addison-Wesley, Reading, MA, 1976.

\bibitem[\protect\citefmt{Schmalzing \bgroup et al.\egroup
  }{1995}]{schmalzing:minkowski}
Schmalzing, J., Kerscher, M., \& Buchert, T., in: \emph{Proceedings of the
  international school of physics Enrico Fermi. Course CXXXII: Dark matter in
  the universe} (Bonometto, S., Primack, J., \& Provenzale, A., eds.),
  Societ\'a Italiana di Fisica, 1995.

\bibitem[\protect\citefmt{Schmalzing}{1996}]{schmalzing:diplom}
Schmalzing, J., Di\-plom\-ar\-beit,
  Lud\-wig--Ma\-xi\-mi\-li\-ans--Uni\-ver\-si\-t{\"a}t M{\"u}n\-chen, 1996, in
  German, English excerpts available.

\bibitem[\protect\citefmt{Schneider}{1993}]{schneider:brunn}
Schneider, R., \emph{Convex bodies: the Brunn-Minkowski theory}, Cambridge
  University Press, Cambridge, 1993.

\bibitem[\protect\citefmt{Seto \bgroup et al.\egroup }{1997}]{seto:nonlinear}
Seto, N., Yokoyama, J., Matsubara, T., \& Siino, M., Ap. J. Suppl. \textbf{110}
  (1997).

\bibitem[\protect\citefmt{{Smoot} \bgroup et al.\egroup
  }{1994}]{smoot:topologyfirst}
{Smoot}, G.~F., {Tenorio}, L., {Banday}, A.~J., {Kogut}, A., {Wright}, E.~L.,
  {Hinshaw}, G., \& {Bennett}, C.~L., Ap. J. \textbf{437} (1994), 1--11.

\bibitem[\protect\citefmt{Stratonovich}{1963}]{stratonovich:topicsI}
Stratonovich, R.~L., \emph{Topics in the theory of random noise}, Vol.~1,
  Gordon and Breach, New York, 1963.

\bibitem[\protect\citefmt{ter Haar~Romeny \bgroup et al.\egroup
  }{1991}]{koenderink:scalespace}
ter Haar~Romeny, B.~M., Florack, L.~M.~J., Koenderink, J.~J., \& Viergever,
  M.~A., in: \emph{Lecture Notes in Computer Science}, Vol. 511, Springer
  Verlag, Berlin, 1991, pp.~239--255.

\bibitem[\protect\citefmt{Tomita}{1990}]{tomita:statistics}
Tomita, H., in: \emph{Formation, dynamics and statistics of patterns}
  (Kawasaki, K., Suzuki, M., \& Onuki, A., eds.), Vol.~1, World Scientific,
  1990, pp.~113--157.

\bibitem[\protect\citefmt{Torres \bgroup et al.\egroup }{1995}]{torres:genus}
Torres, S., Cay\'on, L., {Mart\'{\i}nez-Gonz\'alez}, E., \& Sanz, J.~L., Mon.
  Not. R. Astron. Soc. \textbf{274} (1995), 853--857.

\bibitem[\protect\citefmt{Weil}{1983}]{weil:stereology}
Weil, W., in: \emph{Convexity and its applications} (Gruber, P.~M. \& Wills,
  J.~M., eds.), Birkh{\"a}user, Basel, 1983, pp.~360--412.

\bibitem[\protect\citefmt{Weinberg \bgroup et al.\egroup
  }{1987}]{weinberg:topologyI}
Weinberg, D.~H., {Gott III}, J.~R., \& Melott, A.~L., Ap. J. \textbf{321}
  (1987), 2--27.

\bibitem[\protect\citefmt{White}{1979}]{white:hierarchy}
White, S.~D.~M., Mon. Not. R. Astron. Soc. \textbf{186} (1979), 145--154.

\end{thebibliography}
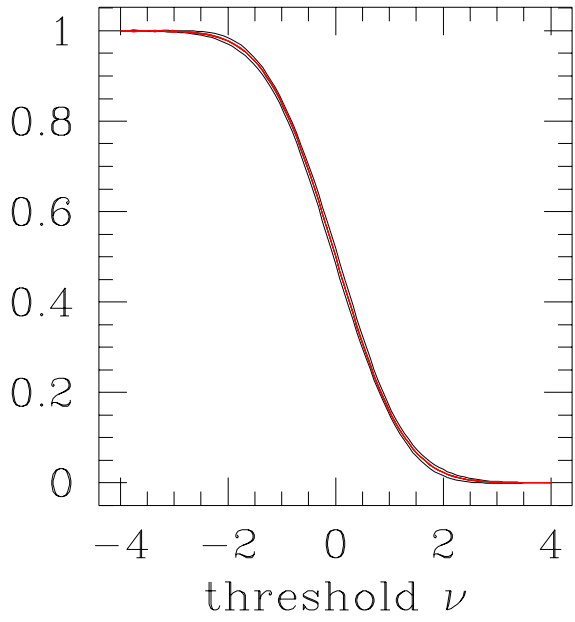
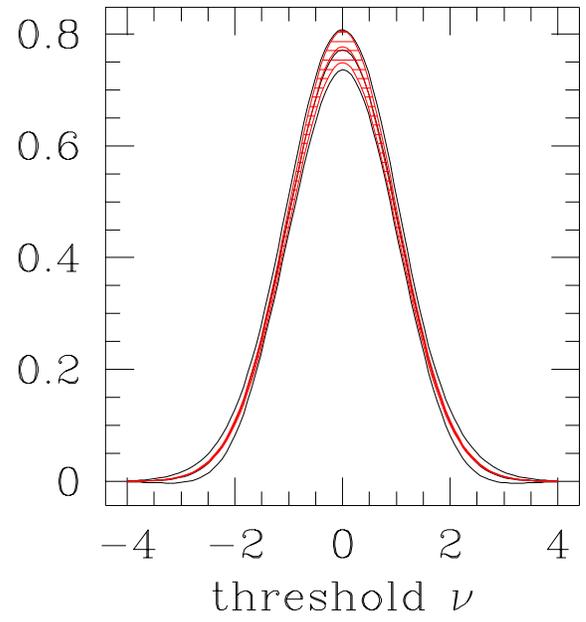
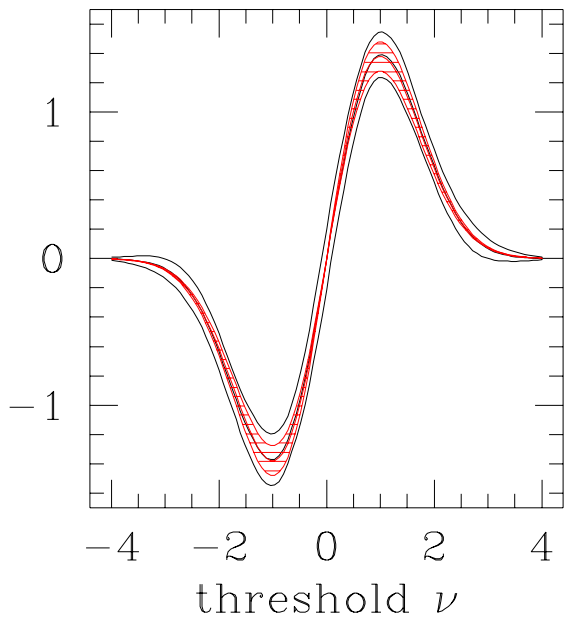
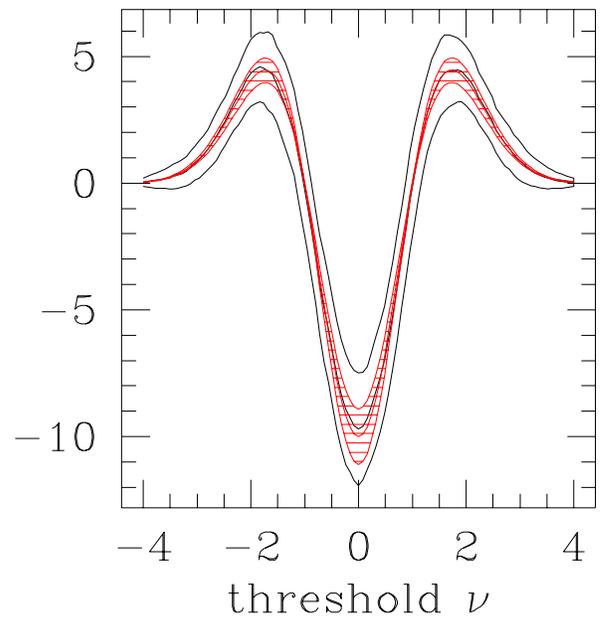
\end{document}